\documentclass[twocolumn,twoside,slac_two]{revtex4}
\usepackage{graphicx}
\usepackage{fancyhdr}
\usepackage{hyperref}
\hypersetup{
    colorlinks=true,
    urlcolor=magenta,
}
\begin{document}

\title{Detection of cosmic rays in the PeV to EeV energy range}

\author{Donghwa Kang}

\affiliation{
  Institut f\"ur Kernphysik,
  Karlsruher Institut f\"ur Technologie,
  76021 Karlsruhe, Germany
}

\begin{abstract}
Cosmic rays around the knee are generally believed to be of galactic origin. Observations on their energy spectrum and chemical composition are important for understanding the acceleration and propagation of these cosmic rays. In addition, it is required to clarify the transition from galactic to extragalactic sources. In this paper, results of recent experiments measuring around the knee will be reviewed along with the detection techniques. The results on the all-particle energy spectrum and composition in the energy range of the knee up to the ankle will be discussed.
\end{abstract}

\maketitle

\thispagestyle{fancy}

\section{Introduction}
Since the discovery of cosmic rays more than a century ago,
many investigations of high-energy cosmic rays have been performed 
to find their sources.
However, the sources and acceleration mechanism are still unknown. 
The aims of experimental cosmic-ray studies are to determine 
the arrival direction, the energy spectrum and chemical composition of cosmic rays.
Those studies and its results are fundamental to understand their origin,
acceleration and propagation.

In general, the all-particle energy spectrum of cosmic rays 
follows a power law $dN/dE$ $\propto$ $E^{\gamma}$ 
with a spectral index of $\gamma$ around $-3$.
Particles up to the energy of about 10$^{14}$\ eV can be measured directly by
balloon or satellite experiments, whereas for higher energies direct
measurements cannot provide data with sufficient statistics due to their small
sensitive detection area and exposure time. Experiments thus have to observe
cosmic rays indirectly by measuring extensive air showers.

Around 10$^{15}$\ eV, the all-particle spectrum has a power-law-like 
behavior with $\gamma \sim -2.7$.
The prominent feature is known as the knee at $3-5 \times 10^{15}$\ eV, 
where the spectral index changes from about $-2.7$ to $-3.1$.
The general explanation of the steepening of the spectrum is 
due to the reach of the maximum energy of galactic acceleration mechanisms 
of the cosmic rays \cite{Hillas} or due to propagation effects,
so that the energy positions of the knees for different
cosmic ray primaries would be expected to depend on their atomic number,
i.e. the charge.

Supernova remnants are generally believed to be the sources for cosmic rays
from about 10\ TeV up to PeV with acceleration of the particle by the
first order Fermi mechanism. The cosmic ray particles gain their
final energy by many interactions each with a small increase of the energy
emitted by supernova explosions.
The maximum attainable energy of charged particles is obtained by 
$E_{knee} \propto ZBR$, where $Z$ is the cosmic ray particle charge, 
$B$ and $R$ are the magnetic field strength and the size of the acceleration
region, respectively,
fitting more or less to the maximum energy of the knee.

In the energy range between $10^{16}$\ eV and $10^{19}$\ eV, 
an important feature is the ankle, which is characterized 
by a slight flattening followed by a steepening,
and at about $6\times10^{19}$\ eV by a steep cutoff.
Cosmic rays above the ankle are most probable of extra-galactic origin,
so that in the energy range between knee and ankle
a breakdown of the heavy component and
a transition from a galactic to an extra-galactic dominated composition 
are expected \cite{Hillas, Berezinsky, Haungs}. 

A continuous and steady source would generate the distribution 
of the energy spectrum with a simple power law for all the elements. 
However, if sources are discrete and distributed non-uniformly
in space and time, it might generate structures and changes
in the spectral indices of the individual primaries at certain energies. 
This feature would be more pronounced at higher energies.
Therefore, investigations on the cosmic ray primary spectrum, 
composition and anisotropy are important to address the above questions.

Measurements performed by many experiments
using different observation techniques in the energy interval of
PeV to EeV showed the existence of individual knees 
in the spectra of the light and the intermediate mass groups of cosmic rays.
In this paper, the following three experiments presently
under consideration are discussed:
KASCADE-Grande measures the electromagnetic components with scintillation detectors
and in addition low energy muons of air showers with shielded scintillators.
Tunka-133 measures the Cherenkov light
emitted by the electromagnetic components with wide angle photomultipliers.
IceTop uses the Ice-Cherenkov tanks to measure air showers and
has a possibility to include the detection of high-energy muons with IceCube.

First, a brief description of the three experimental apparatus and
their results on the all-particle energy spectrum
as well as mass composition are discussed.
After that, a short discussion on the implication on the results are followed.

\begin{figure}[t!]
\begin{center}
\includegraphics[width=0.46\textwidth]{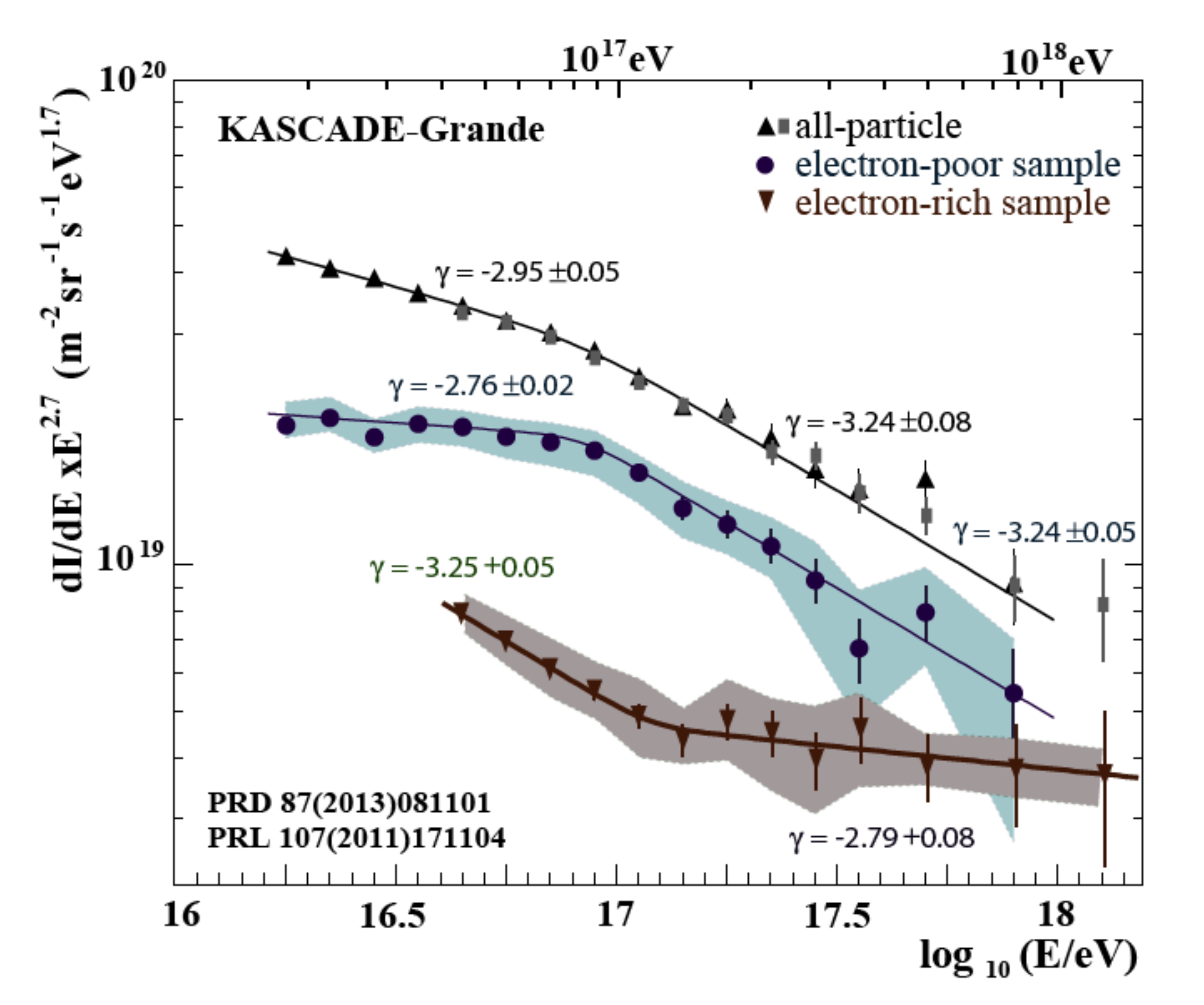}
\caption{The all-particle, heavy-mass enriched, and light-mass enriched energy spectra
  measured by KASCADE-Grande \cite{Apel2, Apel3}.}
\label{fig1}
\end{center}
\end{figure}

\begin{figure}[t!]
\begin{center}
\includegraphics[width=0.46\textwidth]{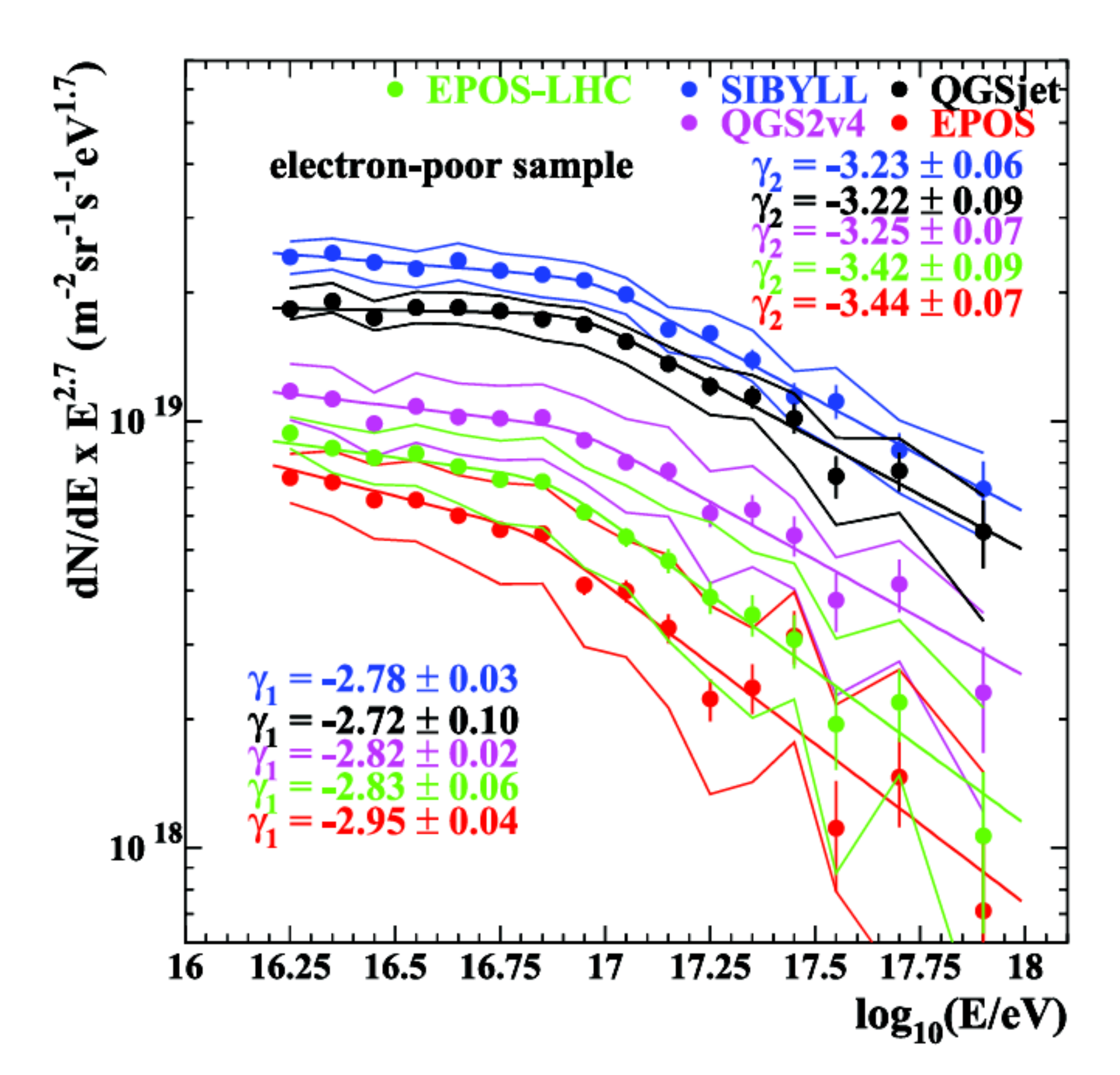}
\caption{
  The energy spectra of heavy-mass enriched events from the KASCADE-Grande data
  based on the different hadronic models of
  QGSJet01, SIBYLL-2.1, QGSJetII, EPOS, and EPOS-LHC \cite{Bertaina}.}
\label{fig2}
\end{center}
\end{figure}

\section{Measurement of air showers}
\subsection{KASCADE-Grande}
The measurement technique at KASCADE-Grande is the large air shower array
with scintillation counters,
while the low energy muons are measured with scintillation detectors underground.

The KASCADE-Grande \cite{Navarra, Haungs2} was an extensive air shower experiment located at
the Karlsruhe Institute of Technology, Karlsruhe, Germany 
(49.1$^{\circ}$ north, 8.4$^{\circ}$ east, 110\ m above sea level). 
The KASCADE-Grande experiment was the extension of the original KASCADE array
designed to measure primary cosmic rays
in the energy range of 10$^{16}$ to 10$^{18}$\ eV.
The main detector components of KASCADE-Grande are the KASCADE array, the Grande
array and the muon detection devices.
The Grande array covering an area of 700$\times$700\ m$^{2}$ extends cosmic
ray measurements up to primary energies of 1\ EeV.
It comprises 37 scintillation detector stations located on a hexagonal grid
with an average spacing of 137\ m for the measurements of the charged shower
component. Each of the detector stations is equipped with plastic scintillator
sheets covering a total area of 10\ m$^{2}$.

The primary energy of cosmic rays are reconstructed by the observed electron
and muon numbers at ground. While the Grande detectors are sensitive to all
charged particles, the KASCADE detectors measure separately the
electromagnetic and muonic components due to the shielding above the muon
counters. Therefore, the shower core position, the arrival direction, and the
total number of charged particles in the shower are reconstructed from Grande
array data, whereas the total number of muons is extracted from the data of
the KASCADE muon detectors. 

The estimation of the energy and mass of the primary cosmic rays
is performed by a combined investigation of the charged particles
and the muon components
measured by the detector arrays of Grande and KASCADE.
In KASCADE-Grande, the two-dimensional shower size distribution of
charged particle number ($N_{ch}$) and muon number ($N_{\mu}$) 
is the basis for the determination of energy and mass. 
Using the hadronic interaction models of QGSJet-II,
the all-particle energy spectrum was determined
in the energy range of $10^{16} - 10^{18}$ eV
with a total uncertainty in flux of 10\ \%. 
In the resulting all-particle energy spectrum \cite{Apel1},
some structures are observed, not described by a single power law:
a concave behavior just above $10^{16}$ eV and a small break,
i.e. a knee-like feature at around $10^{17}$ eV.
This knee-like feature occurs at the energy,
where the knee of the heavy primaries, mainly iron, would be expected.

The reconstruction of the energy spectra for individual mass groups
is performed by subdividing the measured data into two samples,
which are defined as heavy and light mass groups \cite{Apel2, Apel3}.
The selection for different mass groups
is based on the correlation between the size of $N_{ch}$ and $N_{\mu}$
on an event-by-event basis, the so-called $k$-parameter.
Using the separation parameter $k$ determined by simulations
the energy spectra of the mass groups are obtained in Fig. 1.
For this result, the air shower simulation was done
by the CORSIKA package \cite{Heck}
with the hadronic interaction model of QGSJet-II.
The results show that a clear knee-like feature
in the electron-poor spectrum,
i.e. heavy elements of primary cosmic rays, is observed
at an energy around $8\times10^{16}$ eV
with a change of index $\Delta\gamma \sim -$0.48.
This slope change is much more significant than the one
at the all-particle spectrum with $\Delta\gamma \sim -$0.29.
In addition, in the electron-rich spectrum,
an ankle-like structure was observed at an energy of $10^{17.08}$ eV
with the change of the spectral index of $\Delta\gamma \sim -$0.46.

The energy calibration of air shower events depends on hadronic interaction models,
which describe the development of the extensive air showers in the atmosphere.
Figure 2 shows the resulting reconstructed energy spectra of the heavy primaries
on basis of energy calibration
with five different high-energy hadronic interaction models.
Even though different hadronic interaction models predict
some different abundances of the elemental groups,
the shape and the structure of the obtained energy spectra
for the heavy and light primaries of cosmic rays remain unchanged
\cite{Bertaina, Kang}.

Results from KASCADE-Grande have observed additionally two more features: 
a knee-like structure in the heavy primary spectrum at around 80 PeV
and an ankle-like structure in the light primary spectrum at an energy of 100 PeV.

An additional confirmation of these features
comes from the combined analysis \cite{Schoo1, Schoo2}
of KASCADE and KASCADE-Grande array
with an extension of the fiducial areas shown in Fig. 3. 
The shower reconstruction is done from the events
measured by both KASCADE and KASCADE-Grade arrays.
Therefore, the additional stations at larger distance allow us
to reach higher energies for KASCADE located events
and higher accuracy for Grande located events.
It also allows measurements to be made over a wide energy range
from 10$^{15}$ up to 10$^{18}$\ eV.
Above all, the main goal for combining analysis is
to obtain the all-particle and mass group spectra
by one consistent method of reconstruction procedure.
\begin{figure}[t!]
\begin{center}
  \includegraphics[width=0.45\textwidth]{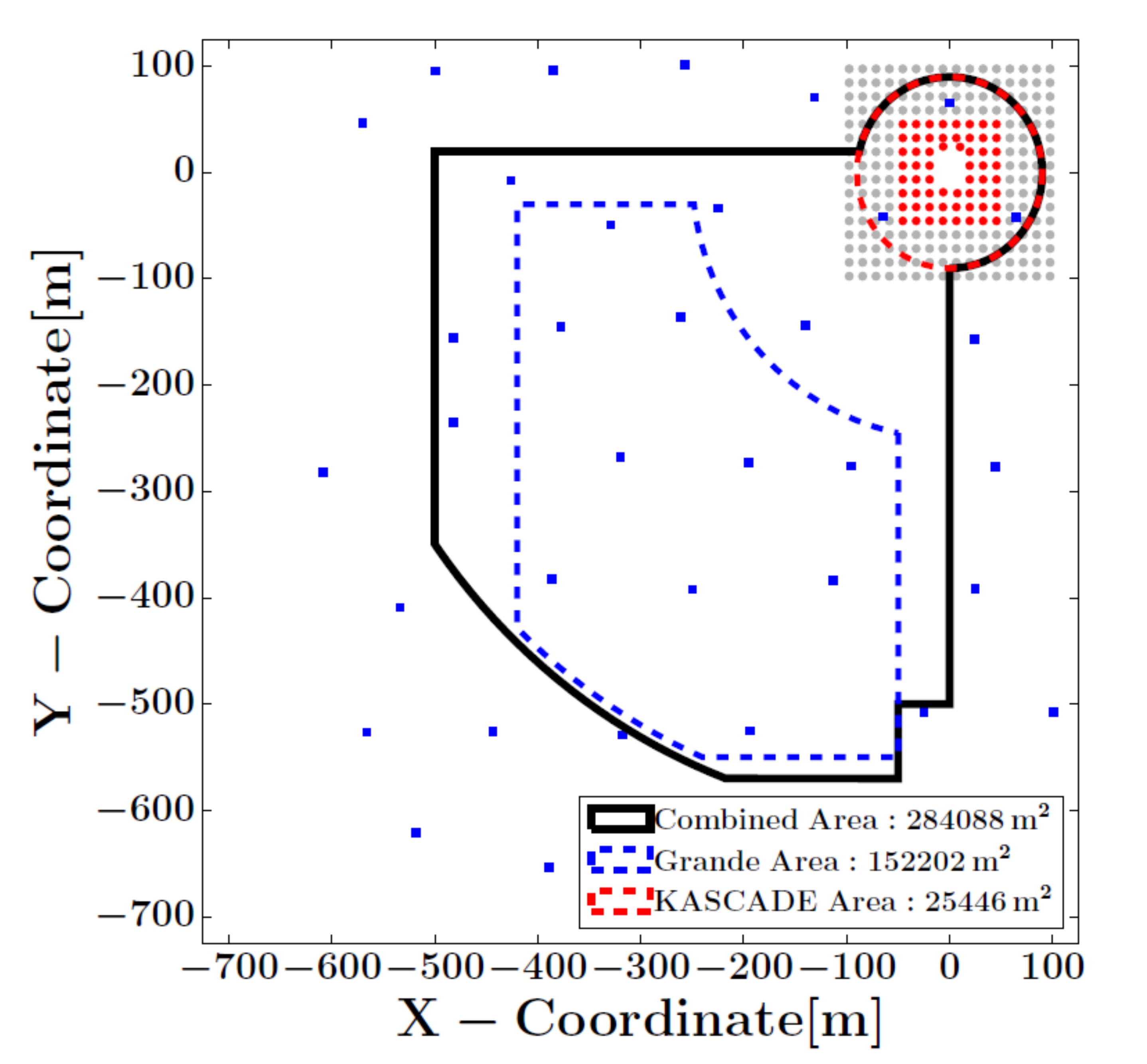}
\caption{
  Schematic view of the KASCADE and KASCADE-Grande experiments.
  The shaded area is covered by 622 m$^{2}$ of muon detectors.
  The KASCADE central detector and muon tunnel are also shown,
  but not used in the combined analysis.}
\label{fig3}
\end{center}
\end{figure}

Figure 4 shows the 2-dimensional shower size spectrum
obtained by the combined shower reconstruction from the data of both arrays.
This distribution from a larger fiducial area is reconstructed
with higher accuracy
and covers more than three orders of magnitude in the primary energy. 
\begin{figure}[t!]
\begin{center}
\includegraphics[width=0.5\textwidth]{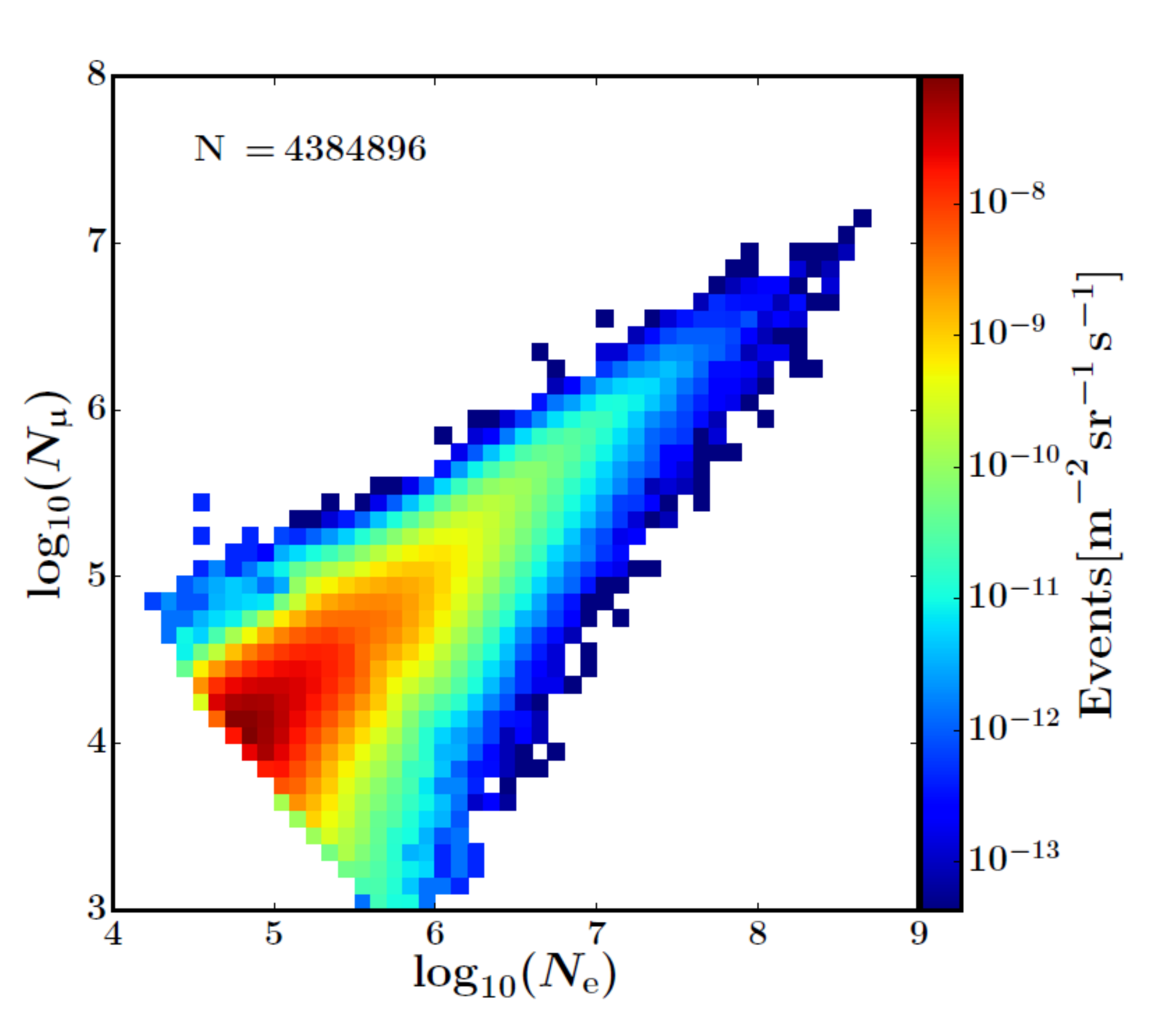}
\caption{
  The 2-dimensional shower size distribution
  for the combined analysis obtained
  by a combined shower reconstruction of
  KASCADE and KASCADE-Grande data \cite{Schoo2}.}
\label{fig4}
\end{center}
\end{figure}

For the combined analysis, the $k$-parameter method is applied
to the combined reconstructed 2-dimensional shower size spectrum
to obtain the all-particle energy spectrum.
An obtained energy spectrum is shown in Fig. 5,
based on the interaction models of QGSJet-II-04 (marks)
and EPOS-LHC (dashed lines).
Considering the spectra based on the QGSJet confirm
the earlier observation of all features:
the light and heavy knees at around 3 and 100 PeV,
and the hardening and the light-ankle at about 10 PeV and 100 PeV, respectively. 

For the comparison of QGSJet-II and EPOS-LHC,
the spectra of light components (H+He) agree quite well,
however, not for heavy generated spectra (C+Si+Fe).
It implies that the proton-proton interaction is better
described in the post-LHC models
than the nucleus-nucleus interaction.
An additionally possible reason for the discrepancy of the heavy spectra is that 
the muon components are not sufficiently described,
since the distance from the shower core covered by muon detectors is limited.
\begin{figure}[t!]
\begin{center}
\includegraphics[width=0.5\textwidth]{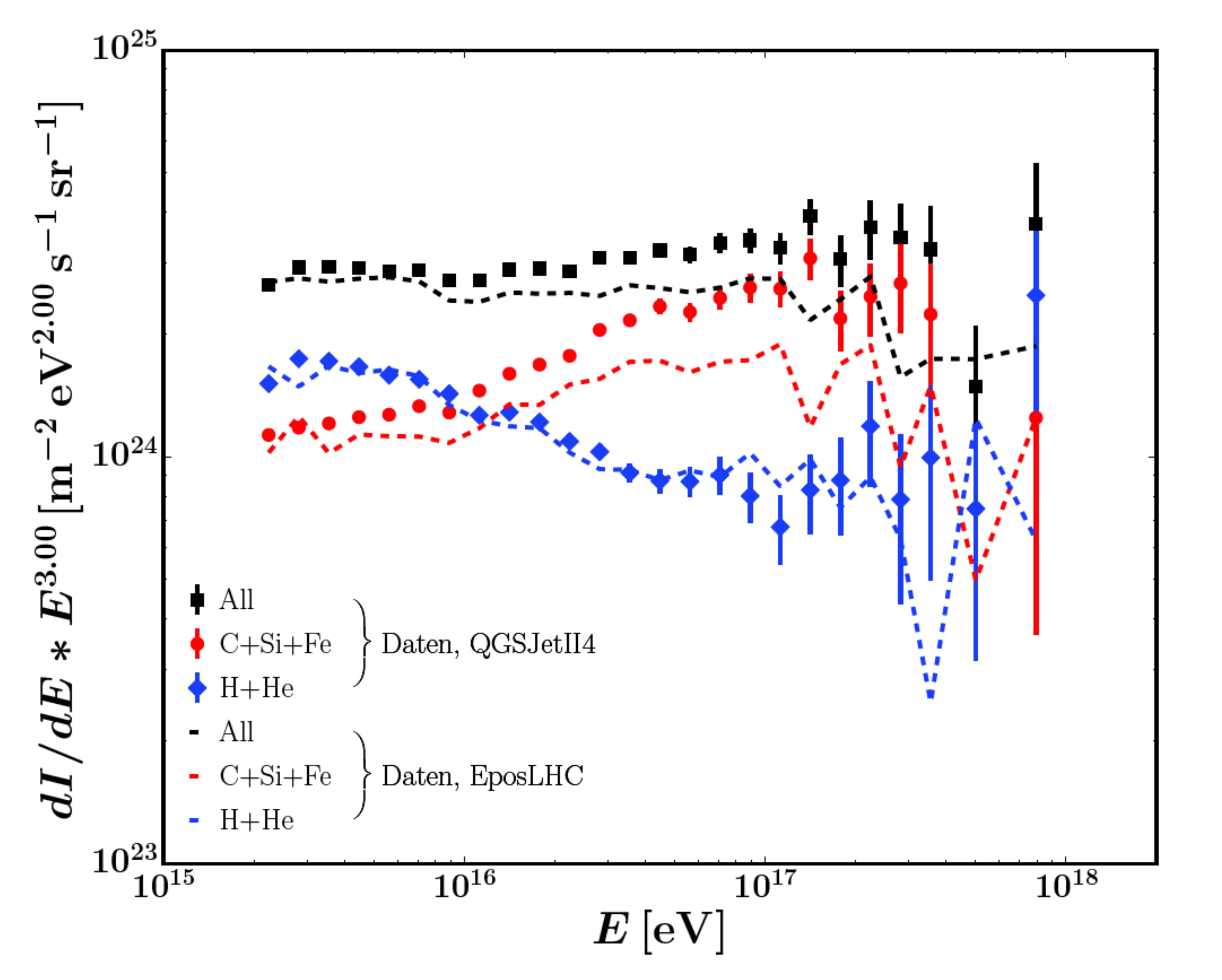}
\caption{
  Reconstructed energy spectra from the combined analysis
  for the heavy and light components using energy calibrations
  based on QGSJet-II-04 (marks) and EPOS-LHC (dashed lines)
  interaction models \cite{Schoo2}.}
\label{fig5}
\end{center}
\end{figure}

\subsection{Tunka-133}
The Tunka-133 experiment \cite{Prosin1} is a Cherenkov light array
located in Tunka Valley in Siberia
and measures the Cherenkov light emitted by extensive air showers
in the energy range from 100 TeV up to 1 EeV.
It consists of 133 non-imaging wide-angle Cherenkov light detectors,
covering an area of 1 km$^{2}$.
Each has a 20 cm diameter photomultiplier.

The arrival direction of air showers are reconstructed
by fitting the measured pulse front delay times using a curved shower front.
The flux density of the Cherenkov light of extensive air showers
is proportional to the primary energy,
so that the primary energy of air showers is determined
from the density of the Cherenkov light flux
at a distance of 200 m, i.e. $Q(200)$ from the shower core.
The relation between the shower energy and the light flux
expressed by $E_{0} = C \cdot Q(200)^{g}$
was derived from the CORSIKA program,
where a mixed composition consisting of equal distribution of proton and iron nuclei
is assumed with the value of the index $g = 0.94$.

The energy spectrum obtained by Tunka-133 is shown in Fig. 6
along with the preliminary spectrum of Tunka-HiSCORE,
which is an extension of measurements to the lower energy range. 
A broken power law fit is indicated in the Figure as well.
The spectral index changes of the spectrum are observed at an energy
about $2\times10^{16}$ eV and $2\times10^{17}$ eV, respectively.
The spectrum structure is similar to the one measured by KASCADE-Grande and
two spectra obtained by different measurement technique are well in good agreement,
by applying a weighting factor of 7\% to the overall flux \cite{Prosin2}. 

The depth of the shower maximum $X_{max}$ is determined from two parameters:
the pulse width at a shower core distance of 400 m
and the steepness of the amplitude distance function. 
The dependence of the mean $X_{max}$ as a function of the primary energy
in the energy range of $10^{16}$ to $10^{18}$eV is presented in Fig. 7.
Tunka-133 results for 5 years measurements \cite{Prosin3} are compared
with the points of the HIRES-MIA experiment
and fluorescent light detectors of the Pierre Auger Observatory (PAO).
In 2015 the Auger expanded the energy range of measurements
with the results of HEAT fluorescent detectors 
and reanalyzed the results of the main detectors
claiming the shift of their points to about 20 g/cm$^{2}$ deeper into the atmosphere.
One can see an agreement of the Cerenkov light Tunka experiment
results with the previous fluorescent light observations.
However, not enough statistics of Tunka-133 is available to discuss
the discrepancy with the current PAO results.

Tunka-133 has a core accuracy of the shower reconstruction less than 10\ m
and an energy resolution of about 15\% for the individual events.
The accuracy of $X_{max}$ is smaller than 25\ g/cm$^{2}$
and the angular resolution is less than 0.3$^{\circ}$.

\begin{figure}[t!]
\begin{center}
  \includegraphics[width=0.45\textwidth]{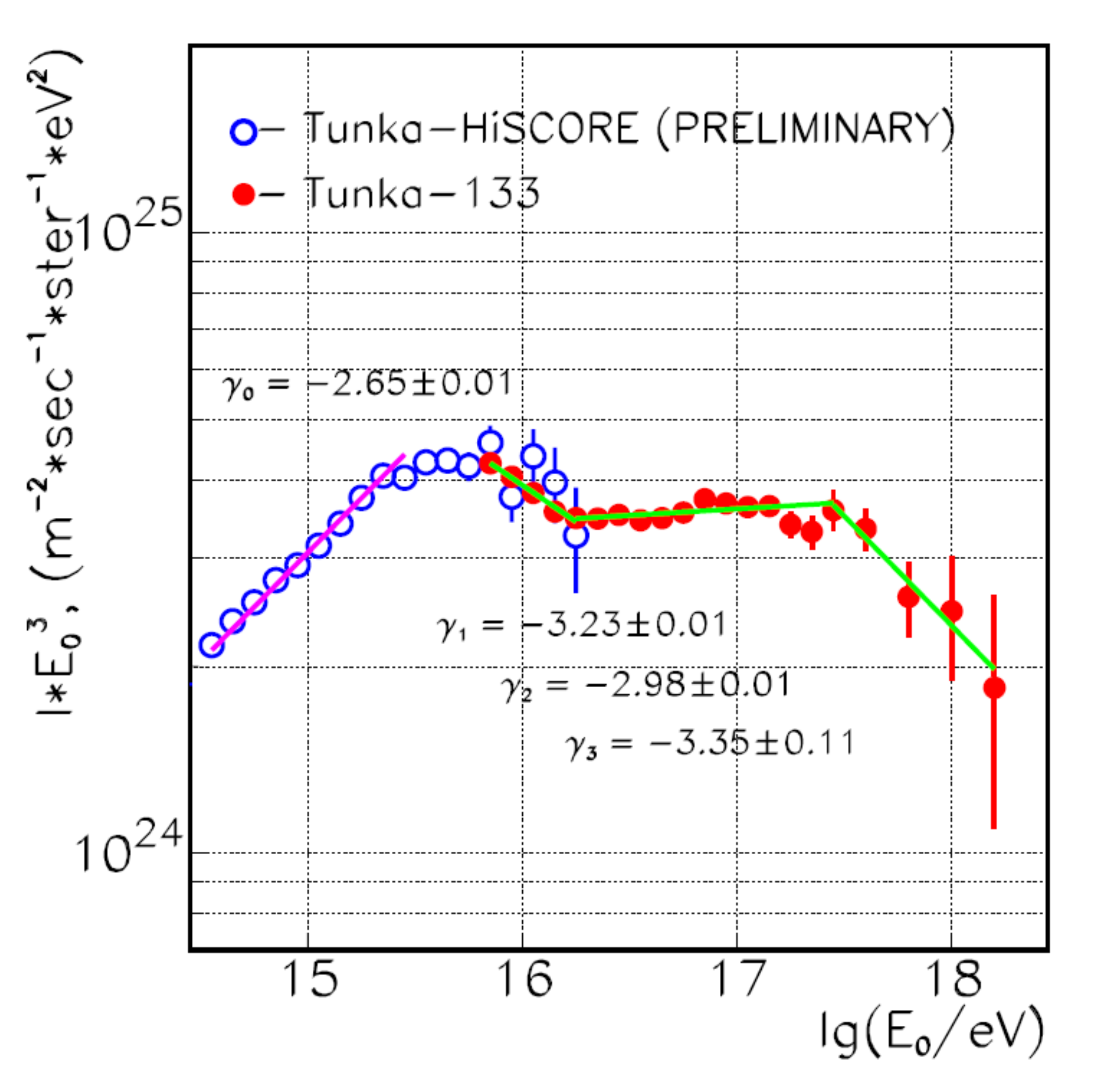}
\caption{
  The all-particle energy spectra measured by Tunka-133 (red circles)
  along with the preliminary spectrum of
  Tunka-HiSCORE (blue open circles) \cite{Prosin2}.}
\label{fig6}
\end{center}
\end{figure}

\begin{figure}[t!]
\begin{center}
\includegraphics[width=0.45\textwidth]{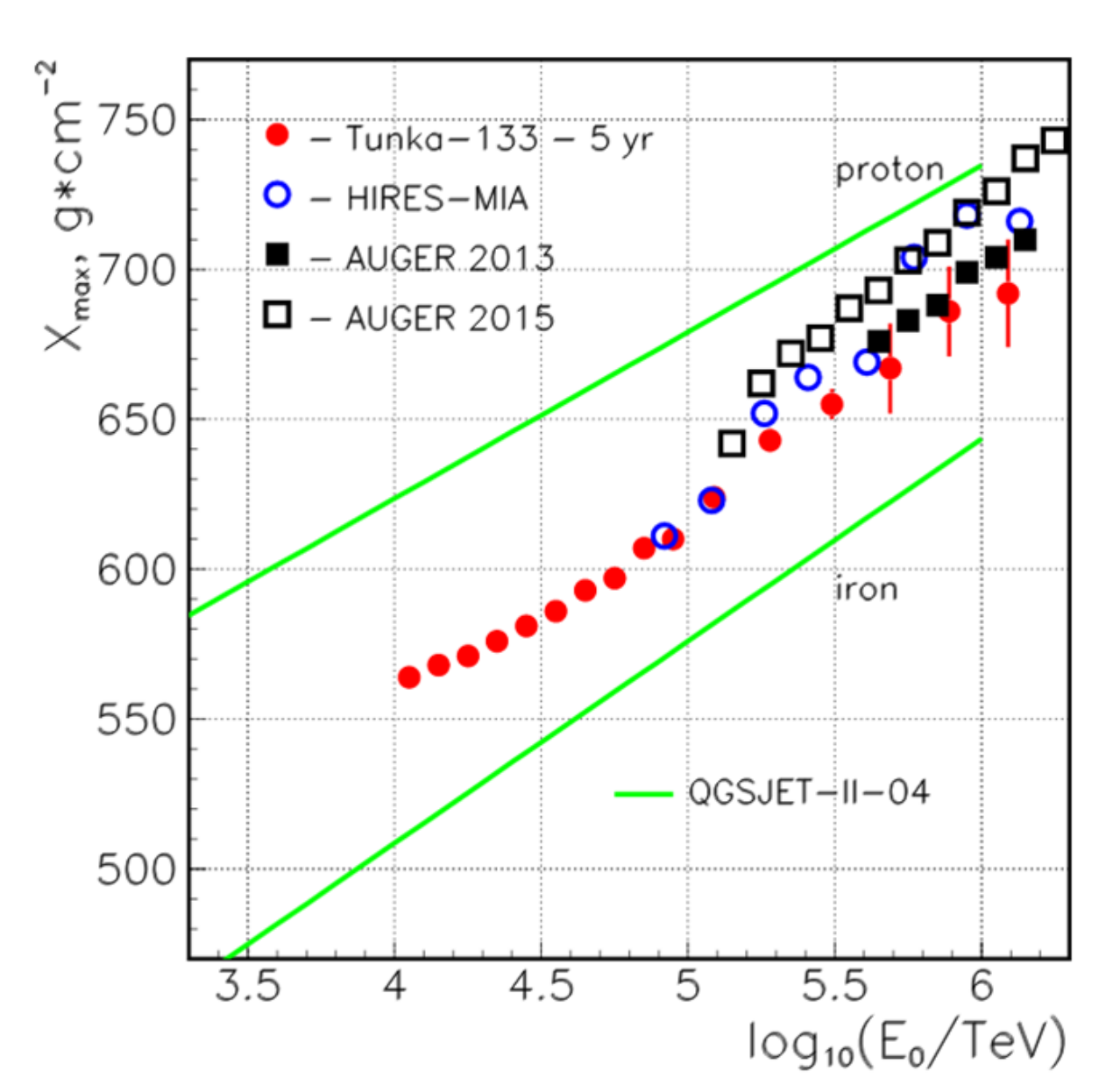}
\caption{
  Dependence of the mean $X_{max}$ measured by Tunka-133
  as a function of energy \cite{Prosin3}.}
\label{fig7}
\end{center}
\end{figure}

\subsection{IceTop}
IceTop is the surface array of the IceCube Neutrino Observatory
at the South Pole \cite{Aartsen, Karg}.
IceTop has an array area of 1 km$^{2}$ with Ice-Cherenkov tanks
and consists of 80 detector stations with each 2 tanks separated
from each other by 10\ m.
It is sensitive to the primary energy of cosmic rays from 1 PeV to 1 EeV.

The IceTop stations detect mainly the signal of the electromagnetic component
of the air shower because of the high-altitude (2835m) observation level.
The air shower is detected by the Cherenkov light in the ice of the tanks,
which is emitted from the charged particles.
Events passed through IceTop and muons measured with different heights
can be reconstructed in both detectors.
The recorded signal in the stations is calibrated
in terms of vertical equivalent muons.
The properties, e.g. shower size, shower direction and core location,
of the primary cosmic ray
are reconstructed by fitting the measured signals
with a lateral distribution function,
which includes an attenuation factor due to the snow cover on top of each tank.
The signal times are fitted with a function describing the shape of the shower front.
The primary energy is then given by the shower size $S_{125}$,
defined as the signal at a lateral distance of 125 m from the shower axis.

IceTop measurements are combined with the signal of high-energy muons
measured with the in-ice IceCube installation
and low-energy muons measured by IceTop arrays at large distances to the shower core
to determine the energy and the elemental composition.

Figure 8 shows the reconstructed all-particle energy spectrum \cite{Rawlins1}
from the events detected only by the surface array IceTop (red marks),
where the conversion of the shower size $S_{125}$ to the energy is performed
assuming a mixed composition by means of the H4a model \cite{Gaisser}. 
In addition, the spectrum in Fig. 8 is obtained from the coincident events
detected by IceTop and the deep-ice detector IceCube (black marks).
A neural network method is used to determine both energy and composition
from the coincident events and the analysis is based
on the hadronic interaction model SIBYLL-2.1.
The snow attenuation calculation and light propagation models are improved.

A good agreement between two spectra is shown and confirmed the knee-like structures:
a smooth change, referred as the knee, between 4 to 7 PeV,
a hardening at around 18 $\pm$ 2 PeV, and a steepening at around 130 $\pm$ 30 PeV.
The significant structures of the spectrum are not
attributed to any of the systematics or detector artifacts.

The mean logarithmic mass $<$ln(A)$>$ for each bin is estimated
by finding the best fit
to measured and simulated data for each energy bin \cite{Rawlins2}.
Figure 9 shows an energy dependence of $<$ln(A)$>$
from the coincident analysis (nominal) and its systematic effects.
It indicates a strong increase in mass up to about 100 PeV,
where the trend changes slope,
however, the systematic uncertainties can largely affect
the measured composition in terms of $<$ln(A)$>$. 
The dominant systematic effect is
from absolute calibration of the light yield in the detector.
A further main uncertainty comes from different hadronic models
similar in scale as the one from the light yield.

\begin{figure}[t!]
\begin{center}
\includegraphics[width=0.5\textwidth]{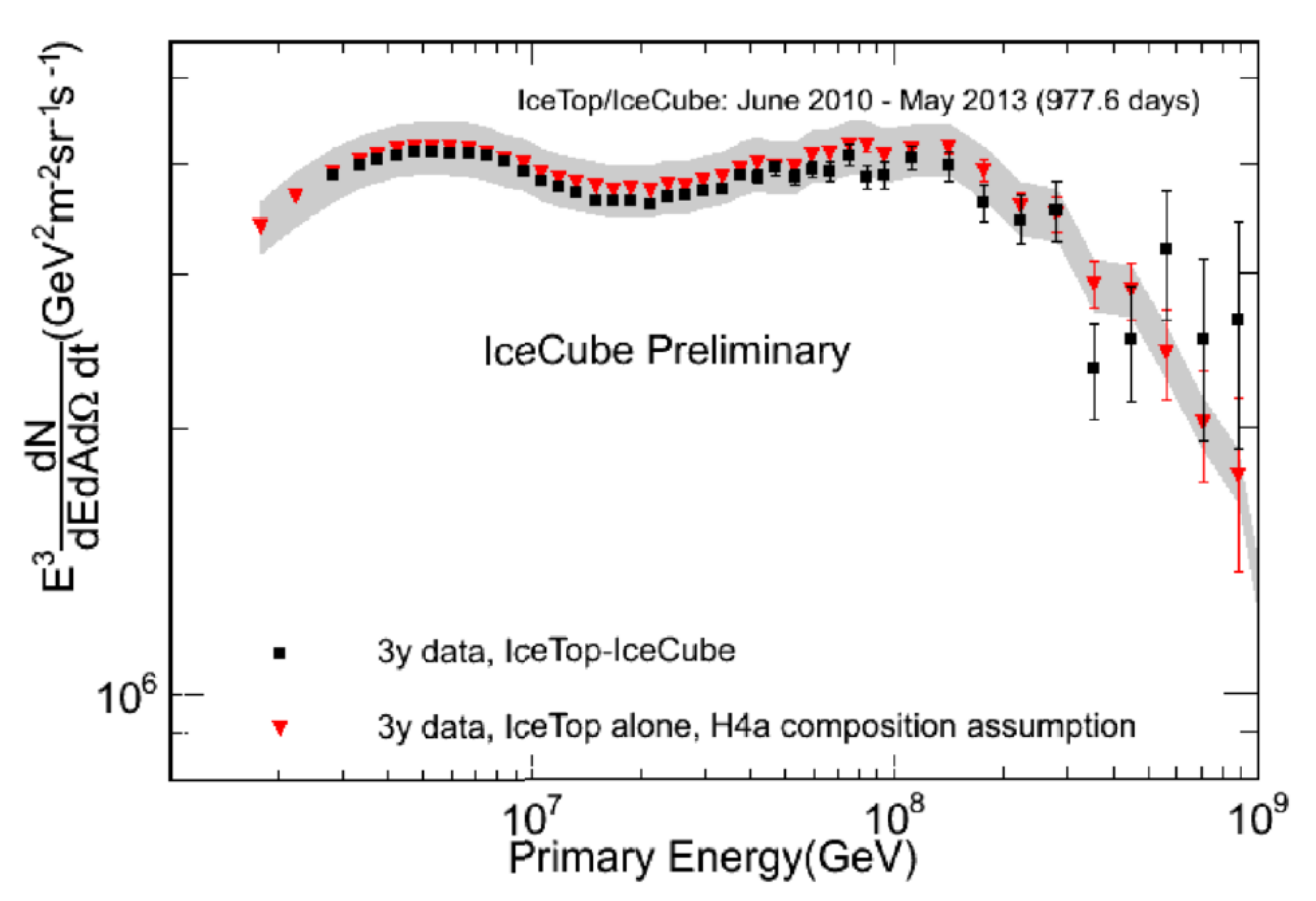}
\caption{
  All-particle energy spectrum obtained from the IceTop three-year data only (black)
  with the spectrum from the coincident analysis (red) \cite{Rawlins2}.}
\label{fig8}
\end{center}
\end{figure}

\begin{figure}[t!]
\begin{center}
\includegraphics[width=0.5\textwidth]{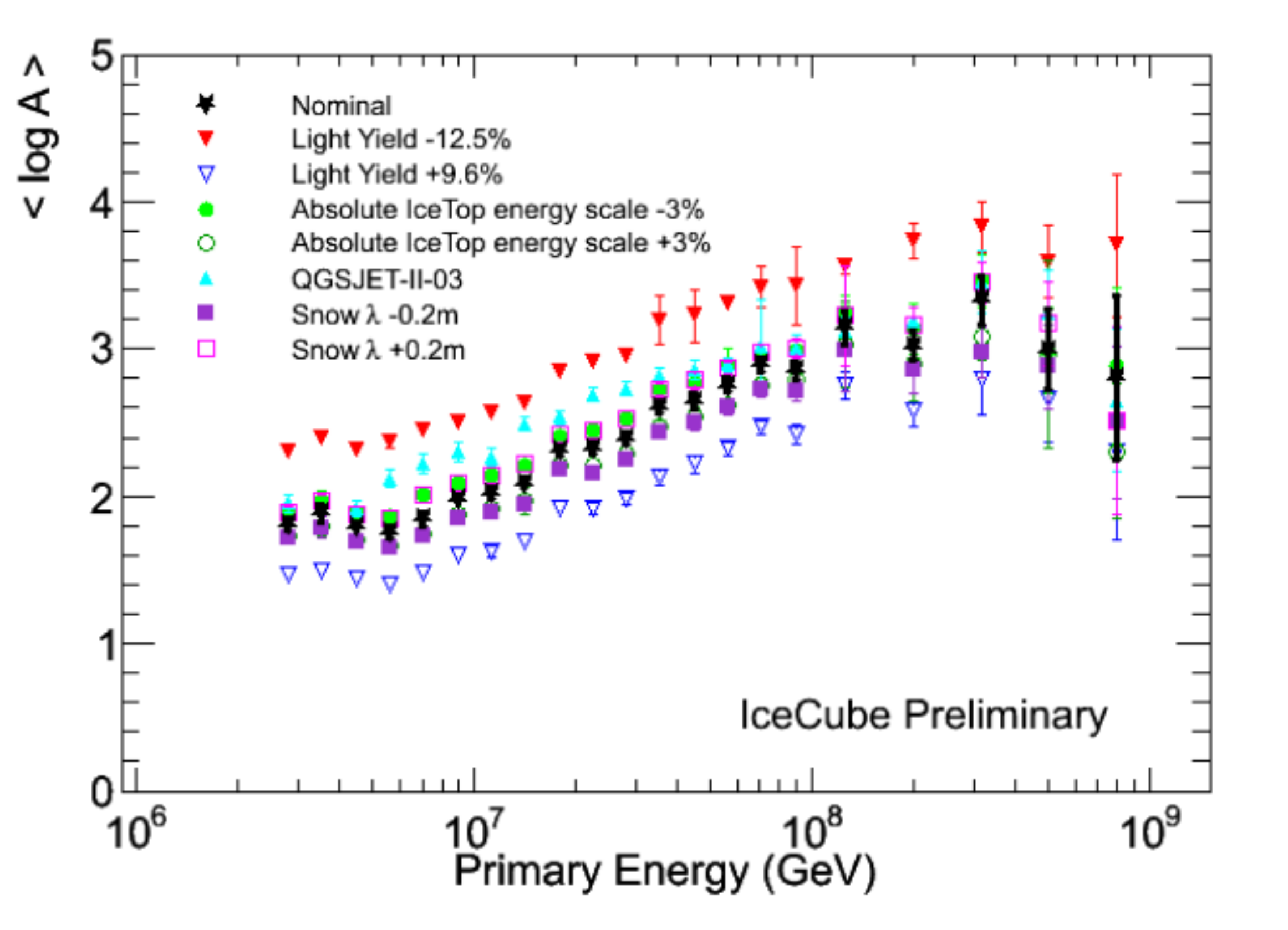}
\caption{
  Energy dependence of the mean logarithmic mass $<$ln(A)$>$
  along with systematic uncertainties.
  The nominal value from the coincident analysis is indicated
  as the star points \cite{Rawlins2}.}
\label{fig9}
\end{center}
\end{figure}

\section{Discussion of all-particle energy spectrum}
Even though many experimental measurements show a smooth power-law feature,
they cannot be described by a single power law.
The well-known structures, e.g. the knee and the ankle, are observed
and in addition there are statistically significant features,
such as a concavity just above $10^{16}$\ eV
and a hardening in the light component above $10^{17}$\ eV.

\subsection{The iron-knee}
Following the rigidity dependence, the position of the knee is predicted to vary from light to heavy elements,
so that the iron represented with the heaviest element is expected to cause a steepening at around 10$^{17}$\ eV 
following previous KASCADE observations \cite{Antoni}.
Therefore, a deeper understanding of the origin of the steepening 
around $10^{17}$\ eV in the spectrum in terms of mass group separation
is interesting and important.
A distinct knee-like feature around $10^{17}$ eV in Fig.\ 10 is observed
by all three experiments
KASCADE-Grande \cite{Apel2}, Tunka-133 \cite{Prosin2}, and IceTop \cite{Rawlins2}
and they support each other.
The position of the spectral break at KASCADE-Grande
is a little lower than by the other two experiments.

Their results are consistent within the order of 15\% of the total flux,
even though three experiments have different measurements technique,
are located at different observation levels,
and use different hadronic interaction models.

A difference between three spectra is the absolute normalization of the energy scale,
which shifts the observed structure slightly in the energy
and also in the absolute flux.
The normalization depends on the calibration and the mass composition,
so the difference occurs due to the assumption of the composition. 
All three spectra show a good agreement within the systematics.
This contribution is taken into account in the systematic uncertainties,
but they do not easily cancel out. 

Measurements of KASCADE-Grande have shown that the spectral structures show very little dependence
on the different hadronic interaction models.
However, the absolute flux has a difference of less than 20\%,
since these differences are related to the
absolute normalization of the energy scale by the various models.
Tunka-133 used a calorimetric method for the energy estimation,
where the energy calibration has a dependence on the hadronic interaction model to a lesser extent.
However, a clarifying explanation by further investigations
of the Tunka data with different hadronic models
is still required.
IceTop results based on the SIBYLL and QGSJet models show a small
difference \cite{Aartsen}.
It might be due to the observation level close to the shower maximum.
In summary, differences between the three experiments
for the same hadronic interaction model
are of the same order of the difference between results
based on different hadronic interaction models at one experiment.

Figure 10 represents the energy spectra of the three experiments
in comparison to other experimental results.
In the overlapping energy range below $10^{16}$ eV,
there is a good agreement with the KASCADE results and others,
although the measurements and data analyses methods are independent.
At higher energy range, IceTop and Tunka-133 results
show a slightly higher flux than the result of KASCADE-Grande,
but they are statistically in agreement with each other and
with other results of the Pierre Auger Observatory and Telescope Array.

\subsection{Concavity}
There is a clear evidence that the spectrum just above 10$^{16}$ eV
shows a concavity which is significant with respect to 
the statistical and systematic uncertainties for all three experiments.
Such a hardening of the spectrum is expected, when a pure rigidity
dependence of the galactic cosmic rays is assumed. 
In this case, 
the gap between light primaries (H and He) and the CNO ($Z = 6-12$) groups 
in their knees requires a hardening of the spectrum \cite{DeDonato}.
However, there are also other astrophysical scenarios possible 
for a concave spectrum. 
%

\subsection{Hardening}
According to recent results,
a knee-like feature in the all-particle energy spectrum of cosmic rays
is observed at around $E = 10^{17}$ eV \cite{Apel2}.
It is due to the steepening in the flux of heavy primaries.
The combined spectrum of light and intermediate mass components 
was found to be compatible with a simple power law.
However, the spectral feature just above 10$^{17}$ eV
shows a change of the slope, namely, a hardening or 
ankle-like feature of light primaries.

In KASCADE-Grande, for such a spectral feature, a more detailed investigation
is performed by means of data with higher statistics.
To obtain increased statistics, a larger fiducial area was used 
and more recent measurements were included.
The selection criteria for the enhancement of light primaries
is optimized as well. 

In the resulting spectrum of the light primaries, a hardening,
i.e. an ankle-like feature is clearly visible \cite{Apel3}.
This might indicate that the transition from galactic to extra-galactic origin
starts already in this energy region.

In astrophysical models,
the transition region from galactic to extra-galactic origin of cosmic
rays is generally expected in the energy range from
10$^{17}$ to 10$^{19}$ eV.
In addition, one should expect a hardening of the proton 
or light primaries components of the cosmic ray spectrum 
to take place below or around 10$^{18}$ eV,
since the onset of the extra-galactic contribution is
dominated by light primaries. 

In general, a transition from one source population to another one 
should result in a hardening of the spectrum.
In this aspect, the KASCADE-Grande result might be the first 
experimental hint to the second galactic component, 
such as the component B proposed by Hillas \cite{Hillas}.
The concavity at about 2$\times$10$^{16}$ eV first claimed by KASCADE-Grande has
been recently confirmed by TUNKA-133 \cite{Prosin2} and IceTop experiments \cite{Rawlins2}.

\begin{figure}[t!]
\begin{center}
  \includegraphics[width=0.5\textwidth]{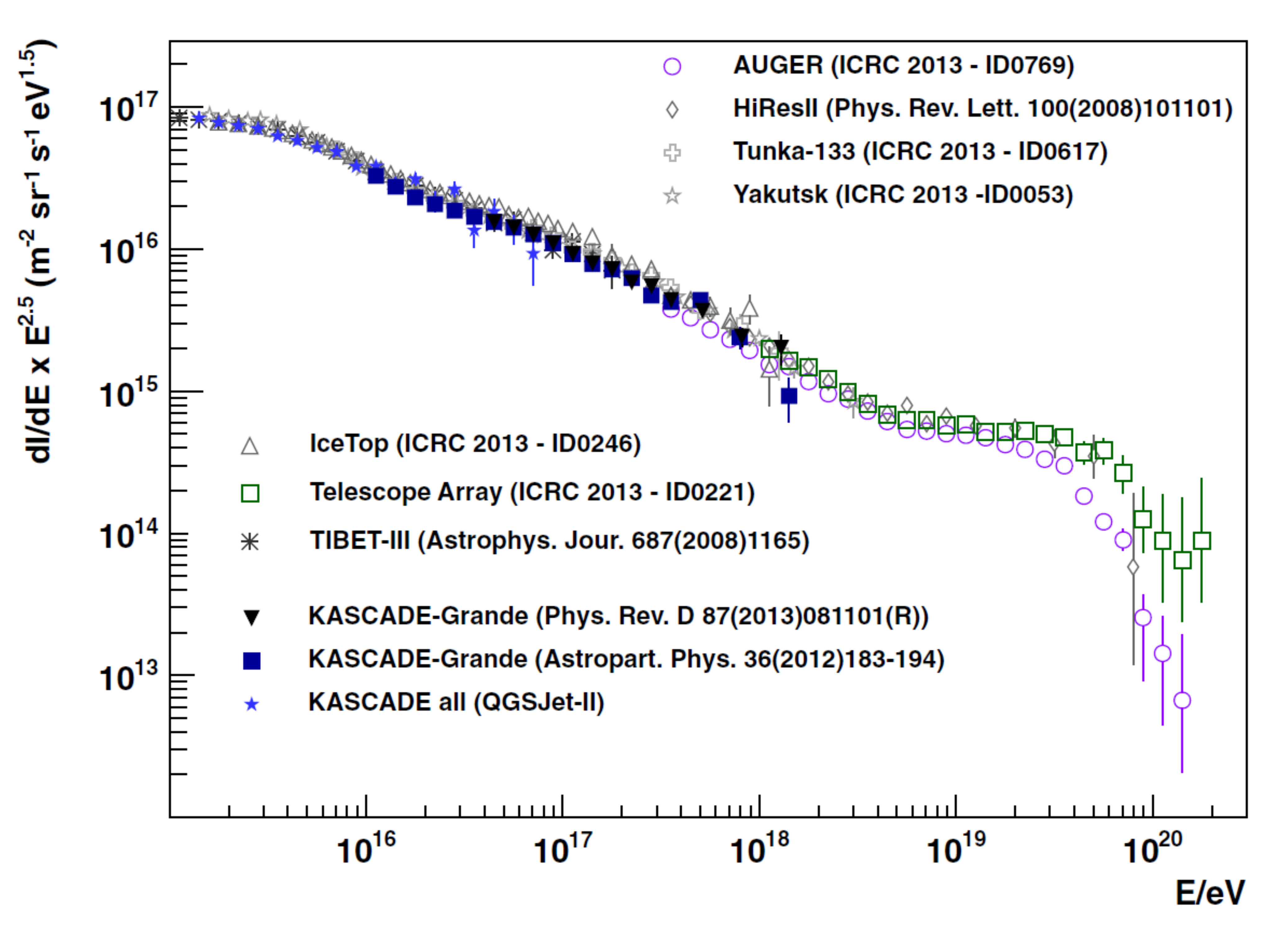}
\caption{
  Comparison of the all-particle energy spectra measured by KASCADE-Grande,
  Tunka-133, and IceTop
  with other experimental results \cite{Haungs3}.}
\label{fig10}
\end{center}
\end{figure}

\section{Discussion of astrophysical implication}
Many experiments indicate that the knee, a sharp steepening,
in the all-particle energy spectrum is caused mainly by a break
in the spectra for the light primaries,
where the mean mass of cosmic rays increases in this region.
However, the interpretation of the knee of the energy spectrum 
of cosmic rays is still under discussion.

One of the most general interpretations for the origin of the knee
is that the bulk of cosmic rays is assumed to be accelerated
in the strong shock fronts of SNRs,
where the spectrum at the source shows a pronounced break. 
The observed knee produced by the steepening of protons
is at an energy of $E_{knee} \approx 4 \times 10^{15}$ eV,
which is possibly close to the size as suggested by SNRs. 

The maximum attainable energy of cosmic ray particles 
has a rigidity dependence \cite{Hillas},
so that the energy position of the knees presents a sequence of steepening 
of different nuclei with increasing $Z$.
The steepening of iron is thus easily expected at an energy of
$26 \times E_{knee}$.
The first evidence for that has been seen by the KASCADE-Grande measurements 
with a knee-like structure of the heavy primary spectrum \cite{Apel2}.
This result seems to support that the structure of the knee
as a rigidity dependent feature.

The acceleration mechanism, i.e. the acceleration of particles 
in $\gamma$-ray bursts is discussed \cite{Wick}.
The $\gamma$-ray bursts associated with supernova explosions
are proposed to accelerate cosmic rays from about $10^{14}$ eV
up to the highest energies.
The propagation effects of cosmic rays is taken into account 
in this approach, and the knee caused by the leakage of particles 
from the galaxy leads to rigidity dependent behavior.

In the model of Hillas \cite{Hillas}, the spectra are reconstructed 
with rigidity dependent knee features at higher energies.
By means of the properties of accelerated cosmic rays in SNRs and
the fluxes derived by KASCADE, Hillas obtained the all-particles flux,
which is insufficient to describe the measured flux
at energy above $10^{16}$ eV.
For this gap, Hillas proposed a second most probable galactic component, 
which is called component B.
An extra-galactic component becomes significant 
at energies above $10^{19}$ eV.
The flux of galactic cosmic rays extends to higher energies in this case,
therefore, a dominated contribution of the extra-galactic component
is expected only above $10^{18}$ eV.

The transition between galactic and extra-galactic cosmic rays occurs
most probably at energies around $10^{17}$ and $10^{18}$ eV.
The transition is an important feature since breaks in all-particle 
energy spectrum and in composition are associated with the particle production
mechanism, the source contribution, as well as their propagation. 

In the model of Berezinsky \cite{Berezinsky}, 
using the model for extra-galactic ultra-high energy cosmic rays 
and the observed all-particle cosmic ray spectrum by Akeno and AGASA,
the galactic spectrum of iron nuclei in the energy range of
$10^{17}$ - $10^{18}$ eV is calculated.
In the transition region of this model, spectra of only galactic iron nuclei
and of extra-galactic protons are present.

The predicted flux
at lower energies is well agreeable with results of the KASCADE data.
The transition from galactic to extra-galactic cosmic rays is obviously
seen in spectra of protons and iron nuclei. Above $10^{17.5}$ eV, 
the spectrum can be described by a proton dominated composition
as is also suggested by $X_{max}$ studies (see Fig.\ 7).

\section{Conclusion}
The all-particle energy spectra of cosmic rays in the PeV to EeV energy range
reconstructed by KASCADE-Grande, Tunka-133, and IceTop 
are mainly discussed in this paper.
The three spectra are well in agreement within systematics,
although they have different observation levels,
different measurement and analysis techniques,
and use different hadronic interaction models
to determine the primary energy spectrum.
Several features have been observed in all three reconstructed energy spectra:
The first dominant feature is
a hardening of the spectrum just above $E = 10^{16}$\ eV
and the main feature is a knee-like feature
in the spectrum of the heavy primaries of cosmic rays,
as well as in the all-particle energy spectrum, at around $E = 10^{17}$\ eV.
Finally, just above $E = 10^{17}$\ eV, an ankle-like structure,
i.e. a remarkable hardening,
in the energy spectrum of light components of cosmic rays
is observed by KASCADE-Grande.
This implies for the first time that the transition 
from galactic to extra-galactic origin of cosmic rays might 
occur already in this energy region.

By means of different hadronic interaction models, there is a shift in the
absolute energy scale of the resulting spectra,
but the shape of the spectrum with its structures remains.
Nevertheless, a reduction of the uncertainty in the hadronic interaction models
used for the shower development is expected.

In addition, the low energy extension of the Pierre Auger Observatory (HEAT)
and Telescope Array (TALE \cite{Ivanov})
will be expected to contribute to high-quality measurements
for the energy range below the ankle in the near future.

The mass composition of KASCADE-Grande, Tunka-133 and IceTop
shows similar tendencies,
however, the absolute scale difference is still large
due to different hadronic interaction models.
Therefore, there is still some uncertainty on the composition around $10^{18}$ eV.

\bigskip 
\begin{acknowledgments}
  The author would like to thank the members
  of the KASCADE-Grande, Tunka-133, and IceTop collaborations.
  In particular, many thanks to Andreas Haungs, Claus Grupen,
  Andrea Chiavassa, Mario Bertaina, and Sven Schoo for helpful discussions and comments.
\end{acknowledgments}

\end{document}